\documentclass[runningheads]{llncs}
\usepackage{graphicx}
\usepackage{xcolor}
\begin{document}
\title{Ensembled Autoencoder Regularization for Multi-Structure Segmentation for Kidney Cancer Treatment}
\titlerunning{Ensemled Segmentation for Kidney Cancer Treatment}
%
\author{David Jozef Hresko\inst{1} \and Marek Kurej\inst{1} Jakub Gazda\inst{2} \and Peter Drotar\inst{1}}
\authorrunning{Hresko et al.}
%
\institute{IISlab, Technical University of Kosice, Kosice, Slovakia \and
2nd Department of Internal Medicine, Pavol Jozef Safarik University and Louis Pasteur University Hospital, Kosice, Slovakia}
\maketitle              
\begin{abstract}
The kidney cancer is one of the most common cancer types. The treatment frequently include surgical intervention. However, surgery is in this case particularly challenging due to regional anatomical relations. Organ delineation can significantly improve surgical planning and execution. In this contribution, we propose ensemble of two fully convolutional networks for segmentation of kidney, tumor, veins and arteries. While SegResNet architecture achieved better performance on tumor, the nnU-Net provided more precise segmentation for kidneys, arteries and veins. So in our proposed approach we combine these two networks, and further boost the performance by mixup augmentation.

\keywords{segmentation \and kidney \and SegResNet \and nnU-Net \and ensemble \and mixup}
\end{abstract}
\section{Introduction}

Kidney cancer (KC) has become one of the ten most common cancer types in the general population, while its incidence steadily increased from the 1970s until the mid-1990s. This trend was attributed to the improved diagnosis secured by the broad introduction of advanced radiological imaging to clinical practice \cite{chow2010epidemiology}. However, KC is strongly associated with risk factors such as cigarette smoking, obesity, and arterial hypertension. Therefore, even though the incidence has recently levelled off, it is not expected to drop anytime soon. On the contrary, it will pose a significant threat in industrialized countries. The treatment of KC is based on ablation, chemotherapy, and surgery \cite{dahle2022renal}, \cite{SHAO2012}. Luckily, most KCs are detected early and only incidentally by radiologic imaging for other diseases when the curative treatment - surgery - is still possible. However, surgery is specifically challenging due to complex regional anatomical relations. Kidney parsing can help tackle this issue and improve preoperative planning and perioperative decisions, leading to a higher chance of successful tumour resection.

Abdominal organ segmentation from medical imaging is a topic that has been researched for several years. Particularly, the computed tomography can provide detailed information about organs and structures in abdominal area. This information can be used for surgical planning, diagnosis and various clinical decisions. In recent years the contour delineation of anatomical and pathological structures became important for applications such as visual support during surgery. Since the manual delineation is tedious and time consuming the automated highly accurate methods are strongly desired.

The advent of convolutional neural networks brought many efficient solution also for  medical images segmentation. The fully convolutional neural network, particularly U-Net architecture \cite{unet2015} became de-facto golden standard for medical image segmentation. Moreover, the pipeline optimisation process implemented in nnU-Net \cite{Isensee2021} boosted segmentation performance and made the U-net easily applicable for any segmentation task. Considering all these recent advances, the abdominal organ segmentation appears to be almost solved problem \cite{abdomenct-1k}. However, even though this may be partially true for a big organs such as liver, kidneys and spleen, the smaller structures still represent challenge for automated segmentation approaches.

Challenges, like LiTS, KiTS, FLARE did not consider smaller structure namely arteries and veins. However segmentation of arteries and veins is crucial for urological surgeries namely laparoscopic nephrectomy \cite{Shao2011},\cite{PORPIGLIA2018}. In this case the precise information about target segmental arteries is required in order to avoid insufficient clamping. 

In this paper we propose ensemble of two encoder-decoder based convolutional neural networks: nnU-net and SegResNet \cite{segresnet}. To further boost the performance of nnU-net, we additionally applied manifold mixup augumentation \cite{manifold}. Similar improvement was already proposed by authors of \cite{kits} on KiTS21 challenge, which improved overall performance of their solution for kidney and kidney tumor segmentation task.

The rest of the paper is organized as follows. In the next section we provide detailed data description. In the methodology section, the proposed solution is explained. Finally we present the results and discuss different aspects of our submission. 


\section{Data}

The data were acquired by Siemens dual-source 64-slice CT scanner and the contrast media was injected during CT image acquisition. The further technical settings of CT are: X-ray tube current is 480 mA, B25F convolution kernel, exposure time equal to 500 ms and voltage 120KV. The slice thickness is 0.5mm/pixel and spacing of images is from 0.47 mm/pixel to 0.74 mm/pixel and 0.75 mm/pixel for z-direction, respectively. 

Altogether there are 130 3D abdominal CT images. The ground-truth corresponding to four classes, kidney, tumor, vein and artery, was determined by three medically trained experts and validated by experienced radiologist. All images were cropped to the same size of $150\times150\times200$ to focus on the four structures of interests 

From 130 cases, 70 are used for training and validation of the model, 30 for open test phase evaluation and 30 for closed test phase evaluation. 

\section{Methodology}

Besides the state-of-art architecture nnU-Net, which is commonly used to solve  medical segmentation tasks, we decided to also adapt and verify not so well-known architectures. To be more concrete, we chose autoencoder based architecture with additional regularization named SegResNet. Based on the obtained results from training phase we decided to ensemble the output from nnU-Net and SegResNet, which were later additionally fine-tuned for closed test phase of KiPA22 challenge.

\subsection{Preprocessing}

In case of the nnU-Net architecture, preprocessing methods including transformation, re-sampling, normalization and scaling were automatically handled by the nnU-Net pipeline. This includes, the resampling strategy for anisotropic data was third order spline interpolation for in-plane and nearest neighbours for out-of-plane. Global dataset percentile clipping along with z-score with global foreground mean and standard deviation was chosen as a normalization strategy. The clipping of HU values were automatically preset with default setting (0.5 and 99.5 percentile).

In case of the SegResNet architecture, as the first step we applied regular channel-wise normalization and scaling to clip CT scan values. After that several augmentations were applied, each with the probability of 0.5. Concretely, we randomly noised original scan with gaussian noise, followed by random rotation, zooming, axis flipping and elastic deformations using bilinear interpolation to calculate output values. 

\subsection{Post-processing}

Based on our observations during training phase, we noticed that some faulty segmentations of kidney and tumor structures were mainly caused by small redundant segments, which were located far from the expected structure location. To prevent this phenomena, we decided to apply keep largest connected component method as the post-processing step during separate inference of both models.

\subsection{Proposed approach}

Our proposed method relies on ensemble technique and consists of two different encoder-decoder based variations of U-Net architecture. The first one is the nnU-Net architecture, which already dominated several segmentation challenges. Following the approach of \cite{kits} we also  took advantege of mixup augmentation. Instead of regular version of mixup \cite{mixup}, we utilized extended version named manifold mixup \cite{manifold}. The key difference here is that, the mixing process of batch samples can be also applied on inner layers of the network, not only on the input layer. The mixed minibatch is then  passed forward regularly from $k$-th inner layer to the output layer, which is  used to compute the loss value and gradients.

\begin{figure}
\centering
\includegraphics[scale=0.6]{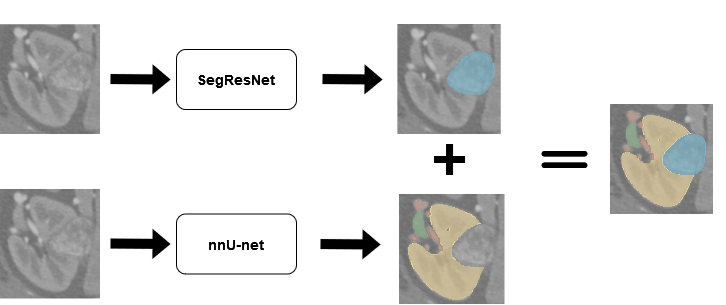}
\caption{Overview of proposed ensemble method} \label{architecture}
\end{figure}

During extensive experiments performed on KiPA22 public dataset we discovered that, there is an architecture, which is capable of better results on some of the segmented classes. Concretely, performance of the nnU-Net was outperformed by SegResNet on tumor class. To benefit from both architectures, we decided to perform ensemble of these two architectures to better localize targeted structures. Based on the obtained results from individual architectures, we keep predictions of kidney, renal vein and renal artery class from nnU-Net, while tumor were produced by SegResNet. Overall concept of proposed method is depicted in Figure \ref{architecture}.

\section{Results}

The evaluation of the model performance is based on two different approaches. The first one relies on the area based metric, concretely Dice Similarity Coefficient (DSC) is used to evaluate the area-based overlap index. The second one is based on distance between predicted and ground truth segmentation mask. Here, to evaluate the coincidence of the surface for less sensitive to outliers, Average Hausdorff Distance (AHD) is used. Additionally, outliers sensitive Hausdorff Distance (HD) is also used to further evaluate the segmentation quality.

To train the final version of nnU-Net, we used stochastic gradient descent (SGD) optimizer with the initial learning rate of 0.01. The length of training was set to 1000 epochs. For manifold mixup augmentation we set the hyperparameter value of mixing coefficient to \textit{$\alpha$} = 0.1. In case of the final version of SegResNet, we used AdamW optimizer with the initial learning rate of 0.0001. The length of training was set to 4000 epochs. For both models we used combined Dice and Cross Entropy loss, batch size equal to two.

The results from the training phase, along with comparisons to other experiments with different models and configurations are presented in Table \ref{table:train}. As can be seen, proposed manifold mixup improved performance of the nnU-Net for each class.  SegResNet outperformed nnU-Net for tumor class segmentation task with notable differences. Based on the \cite{unetr} we were also curious if transformer based architecture is truly capable of better results than standard convolutional network, so we also trained UNETR model on this dataset. Our results proved that this vision transformer was not able to reach the performance of any baseline version of tested models.

\begin{table}[]
\caption{Overall performance of examined models on KiPA22 public train data}\label{table:train}
\begin{tabular}{|c|ccc|ccc|ccc|ccc|l}
\cline{1-13}
\multicolumn{1}{|l|}{} &
  \multicolumn{3}{c|}{Kidney} &
  \multicolumn{3}{c|}{Tumor} &
  \multicolumn{3}{c|}{Vein} &
  \multicolumn{3}{c|}{Artery} &
   \\ \cline{1-13}
Network &
  \multicolumn{1}{c|}{DSC} &
  \multicolumn{1}{c|}{HD} &
  AHD &
  \multicolumn{1}{c|}{DSC} &
  \multicolumn{1}{c|}{HD} &
  AHD &
  \multicolumn{1}{c|}{DSC} &
  \multicolumn{1}{c|}{HD} &
  AHD &
  \multicolumn{1}{c|}{DSC} &
  \multicolumn{1}{c|}{HD} &
  AHD &
   \\ \cline{1-13}
 \begin{tabular}[c]{@{}c@{}}nnU-Net\\(baseline)\end{tabular} &
  \multicolumn{1}{c|}{0.960} &
  \multicolumn{1}{c|}{17.108} &
  0.484 &
  \multicolumn{1}{c|}{0.856} &
  \multicolumn{1}{c|}{12.432} &
  1.412 &
  \multicolumn{1}{c|}{0.805} &
  \multicolumn{1}{c|}{12.002} &
  1.180 &
  \multicolumn{1}{c|}{0.839} &
  \multicolumn{1}{c|}{15.995} &
  0.512 &
    \\ \cline{1-13}
 \begin{tabular}[c]{@{}c@{}}nnU-Net \\ + \\mixup\end{tabular} &
  \multicolumn{1}{c|}{\textbf{0.963}} &
  \multicolumn{1}{c|}{\textbf{17.022}} &
  \textbf{0.424} &
  \multicolumn{1}{c|}{0.893} &
  \multicolumn{1}{c|}{10.114} &
  1.225 &
  \multicolumn{1}{c|}{\textbf{0.823}} &
  \multicolumn{1}{c|}{12.112} &
  \textbf{0.830} &
  \multicolumn{1}{c|}{0.849} &
  \multicolumn{1}{c|}{\textbf{16.395}} &
  \textbf{0.449} &
     \\ \cline{1-13}
UNETR &
  \multicolumn{1}{c|}{0.951} &
  \multicolumn{1}{c|}{19.434} &
  0.874 &
  \multicolumn{1}{c|}{0.838} &
  \multicolumn{1}{c|}{14.768} &
  1.992 &
  \multicolumn{1}{c|}{0.773} &
  \multicolumn{1}{c|}{18.109} &
  3.556 &
  \multicolumn{1}{c|}{0.822} &
  \multicolumn{1}{c|}{20.542} &
  1.987 &
    \\ \cline{1-13}
 \begin{tabular}[c]{@{}c@{}}SegResNet\end{tabular} &
  \multicolumn{1}{c|}{0.961} &
  \multicolumn{1}{c|}{17.045} &
  0.512 &
  \multicolumn{1}{c|}{\textbf{0.901}} &
  \multicolumn{1}{c|}{\textbf{9.100}} &
  \textbf{0.970} &
  \multicolumn{1}{c|}{0.818} &
  \multicolumn{1}{c|}{\textbf{11.224}} &
  1.952 &
  \multicolumn{1}{c|}{\textbf{0.873}} &
  \multicolumn{1}{c|}{19.156} &
 2.443 &
   \\ \cline{1-13}
\end{tabular}
\end{table}

\begin{figure}
\centering
\includegraphics[scale=0.5]{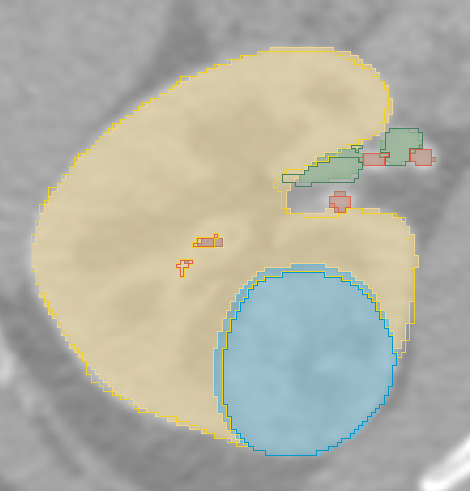}
\caption{Example segmentation for the 61 case. Individual contours denote ground truth label for specific class. Predicted segmentation classes are: yellow=kidney, blue=tumor, green=renal vein, red=renal artery.} \label{good_seg}
\end{figure}

The Fig.\ref{good_seg} shows the example prediction of our proposed method. We randomly selected case 61 from public dataset, which contains all four labels. As can be seen, the segmentation of all structures is almost identical with the ground truth regions. The yellow colour represents healthy kidney tissue located laterally to renal vessels. The red and green colours medially represent the renal artery and renal vein, respectively. Finally, the blue colour dorsally represents kidney cancer. This picture informs the surgeon that the kidney cancer has not invaded renal vessels yet and has a high chance of successful resection. However, a surgeon must evaluate all sections, which may present different spatial relation between kidney cancer and renal vessels. In more complicated cases, human visual evaluation might not be enough to evaluate the ground truth correctly.

Overall performance of our proposed method was also independently measured on KiPA22 challenge test data during open test phase and closed test phase of the challenge. The obtained results from open test phase can be seen in the table \ref{table:opentest}. The results from the closed test phase were not available during the time of writing this paper. 

\begin{table}[]
\caption{Overall performance of proposed method on KiPA22 open test data}\label{table:opentest}
\begin{tabular}{|c|ccc|ccc|ccc|ccc|l}
\cline{1-13}
\multicolumn{1}{|l|}{} &
  \multicolumn{3}{c|}{Kidney} &
  \multicolumn{3}{c|}{Tumor} &
  \multicolumn{3}{c|}{Vein} &
  \multicolumn{3}{c|}{Artery} &
   \\ \cline{1-13}
Network &
  \multicolumn{1}{c|}{DSC} &
  \multicolumn{1}{c|}{HD} &
  AHD &
  \multicolumn{1}{c|}{DSC} &
  \multicolumn{1}{c|}{HD} &
  AHD &
  \multicolumn{1}{c|}{DSC} &
  \multicolumn{1}{c|}{HD} &
  AHD &
  \multicolumn{1}{c|}{DSC} &
  \multicolumn{1}{c|}{HD} &
  AHD &
   \\ \cline{1-13}
  \begin{tabular}[c]{@{}c@{}}Ensembled\\  model\end{tabular} &
  \multicolumn{1}{c|}{0.957} &
  \multicolumn{1}{c|}{17.008} &
  0.464 &
  \multicolumn{1}{c|}{0.880} &
  \multicolumn{1}{c|}{9.104} &
  1.616 &
  \multicolumn{1}{c|}{0.835} &
  \multicolumn{1}{c|}{12.733} &
  0.322 &
  \multicolumn{1}{c|}{0.878} &
  \multicolumn{1}{c|}{16.195} &
  0.322 &
   \\ \cline{1-13}
\end{tabular}
\end{table}

\section{Discussion}

\begin{figure}
\centering
\includegraphics[scale=0.5]{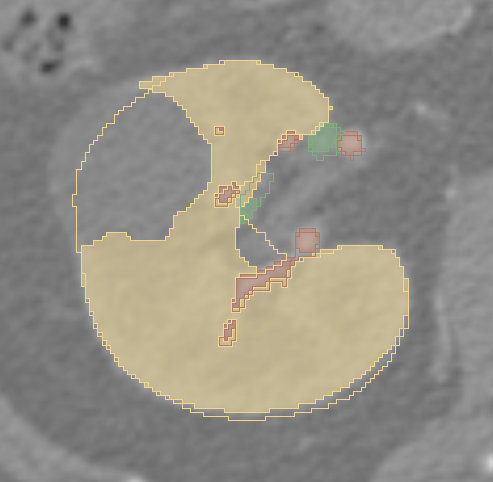}
\caption{Example fault segmentation for the 6 case. Individual contours denote ground truth label for specific class. Predicted segmentation classes are: yellow=kidney, green=renal vein, red=renal artery.}\label{bad_6}
\end{figure}

\begin{figure}
\centering
\includegraphics[scale=0.5]{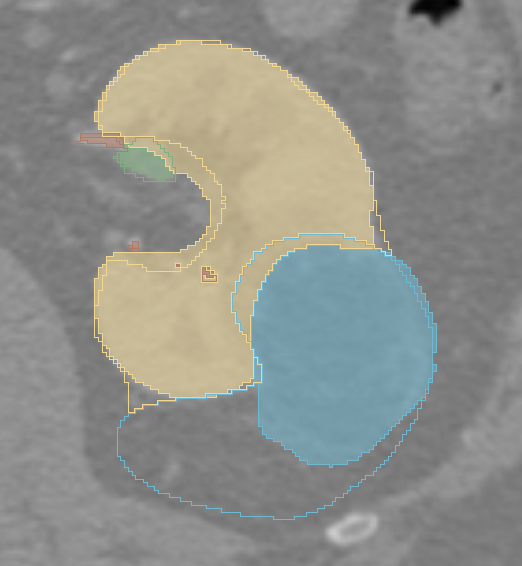}
\caption{Example fault segmentation for the 32 case. Individual contours denote ground truth label for specific class. Predicted segmentation classes are: yellow=kidney, blue=tumor, green=renal vein, red=renal artery.}\label{bad_32}
\end{figure}

Here we investigate in detail some of the segmentation cases that reached lower scores. Fig. \ref{bad_6} represents "a double incorrect case" because neither the ground truth nor the prediction is correct. First, the ground truth (yellow contour, representing healthy renal parenchyma) contains a pathological (probably cystic) lesion (it is impossible to tell without considering other phases of the contrast enhancement). Second, the prediction ignores this lesion and goes around it without rendering and classifying the pathological lesion. Furthermore, Fig. \ref{bad_32} represents "an incorrect prediction case". The ground truth (blue contour) depicts a tumour located in the dorsal part of the left kidney. However, the segmentation fails to render the very dorsal part of the tumour, which again appears as its cystic component. There are no density differences between the cystic component of the tumour and the healthy perirenal fat here, which could be why the prediction failed. Altogether, while it seems that the renal parenchyma, renal vessels and solid tumours have been segmented correctly, the mistakes are associated mostly with cystic lesions or cystic components of kidney cancers.

\section{Conclusions}

In this paper we proposed ensemble method for segmentation of structures in abdominal area. The ensemble is build on combination of nnU-Net and SegResNet architecture. The method achieves highly competitive score on all four structures.

%
%

%
%
%
%
\clearpage
\bibliographystyle{splncs04}
\bibliography{bibtex.bib}
\end{document}